\documentclass[preprint,showpacs,preprintnumbers,amsmath,amssymb]{revtex4}

\usepackage{graphicx}
\usepackage{dcolumn}
\usepackage{bm}

\newcommand{\abs}[1]{\left| #1 \right|}
\newcommand{\sgn}{{\rm sgn}}
\newcommand{\erfc}{{\rm erfc}}

\newcommand{\braketo}[3]{\langle #1 | #2 | #3 \rangle}
\newcommand{\braketi}[2]{\langle #1 | #2 \rangle}

\newcommand{\ket}[1]{| #1 \rangle}

\newcommand{\half}{\frac{1}{2}}

\begin{document}

\preprint{}

\title{Quantum corrections to the masses of the octet and decuplet baryons
in the SU(3) chiral quark soliton model}

\author{Satoru Akiyama}
\author{Yasuhiko Futami}
\email{akiyama@ph.noda.tus.ac.jp}
\affiliation{
	Department of Physics, Tokyo University of Science,
	2641, Noda, Chiba 278-8510, Japan 
}

\date{\today}

\begin{abstract}
Mesonic fluctuations around the chiral solitons are investigated
in the SU(3) chiral quark soliton model.
Since the soliton takes the non-hedgehog shape for the hyperons
and the hedgehog one for the non-hedgehog baryons in our approach,
the fluctuations also change according to the baryonic state.
The quantum corrections to the masses (the Casimir energies) are estimated
for the octet and decuplet baryons.
The lack of the confinement in this model demands the cutoff
on the energy of the fluctuations.
Under the assumption that the value of the cutoff energy is $2\times$(the lightest constituent quark mass),
these calculation reproduces the masses of the baryons within 15 \% error.
\end{abstract}


\maketitle

\section{Introduction\label{sec:intro}}
In the limit of a large number of colors $N_{c}$~\cite{rf:tHooft74},
QCD reduces to an effective theory of the weakly interacting mesons~\cite{rf:tHooft74,rf:Witten79}.
Then the mass of a baryon is proportional to $N_{c}$ 
and the baryons can emerge as solitons of the effective theory~\cite{rf:Witten79}.
It is widely believed that the solitons are important ingredients
of the strong interaction at low energies same as the chiral symmetry.
Some effective models of QCD in low energies region have these feature.
The Skyrme model~\cite{rf:Skyrme61} is a pure mesonic theory and has been investigated
in this context~\cite{rf:Adkins83,rf:Balachandran85,rf:Guadanini84,rf:Mazur84,rf:Callan85,rf:Yabu88}.
The Nambu--Jona-Lasinio (NJL) model~\cite{rf:Nambu61,rf:Ebert86,rf:Weigel92}
and the chiral quark soliton model (CQSM)~\cite{rf:Diakonov88,rf:Reinhardt88,rf:Reinhardt89,rf:Wakamatsu91,rf:Blotz93}
have been studied from a viewpoint of the quark model.
Usually these models are solved by means of the hedgehog ans\"{a}tz and the cranking method
for the non-strange baryons ($N$,$\Delta$).
For the hyperons ($\Lambda$,$\Sigma^{(*)}$,$\Xi^{(*)}$,$\Omega$), in addition,
the embedding of SU(2) meson field to SU(3) one is assumed.

In the previous works~\cite{rf:Akiyama03,rf:Akiyama04}, we investigated the validity of the hedgehog ans\"{a}tz 
for the octet and decuplet baryons in the SU(3) CQSM. 
The isospin vector of the profile function takes the non-hedgehog shape for the hyperons
and the hedgehog one for the non-strange baryon.
And the radial component of the profile function for the hyperons approaches to the center of the soliton
compared with the non-strange baryon.
Thus, the shape of the soliton changes according to the baryon state.

These nature of the soliton is due to the quark mass in the body fixed frame.
The flavor rotation into strange direction rearranges the flavor SU(3) quarks (u,d,s)
and couple the (u,s) or (d,s) quarks as a SU(2) doublet in this frame.
Thus the masses of the doublet are asymmetric under a SU(2) transformation.
That is reasonable, because SU(2) is the ${\rm SU(2)_{V}}$ or ${\rm SU(2)_{U}}$ subgroups of SU(3)
and the asymmetry is a SU(3) symmetry breaking due to the mass difference $(m_{s}-m_{u})$
of s and u quarks.
The non-hedgehog shape of soliton in our approach reflects the asymmetry
and the shrink is due to the strange quark mass. 

The approaches using the hedgehog ans\"{a}tz treat the doublet
as SU(2) symmetric commonly to the octet and decuplet baryons
and incorporate the SU(3) symmetry breaking by a perturbation with respect
to the mass difference $(m_{s}-m_{u})$.
This approach reproduces the mass difference between baryons
but give the too large absolute values.

On the other hand, our approach prepares the doublet and the soliton for every baryon states
and incorporate the SU(3) symmetry breaking at the soliton level in advance.
As a result the Hilbert space of the soliton+quark system is enlarged.
Our procedure reproduces both the absolute values
and the mass differences between hyperons~\cite{rf:Akiyama04}.
However, it gives the large absolute values only for the non-strange baryon masses.
It is because that there is no difference between our approach
and other method using the hedgehog ans\"{a}tz for the non-strange baryons.

For the nucleon mass, there are several attempts to resolve the too large prediction in the soliton models.  
In the Skyrme model, the quantum corrections to the nucleon mass due to the meson fluctuations
around the hedgehog soliton have been evaluated~\cite{rf:Moussallam91,rf:Holzwarth93,rf:Holzwarth94}.
In the context of the quark model, the same corrections are estimated
in the NJL model~\cite{rf:Weigel95,rf:Alkofer95}.
The corrections give large negative contributions to the non-strange baryon masses,
and their results are in good agreement with the experimental values.
The contributions are called the Casimir energies.
Their analyses are valid for the non-strange baryons in our approach.
For the hyperons, however, the non-hedgehog soliton requires some modifications on their procedure.

Thus in this paper we study the Casimir energies due to the meson fluctuations
around the non-hedgehog soliton and check the consistency of our treatment and the Casimir energy
for the octet and decuplet baryons.

In Sec.~\ref{sec:casimir_energy} we show the formal definition of the Casimir energy.
In Sec.~\ref{sec:su3_cqsm} after a brief review of the SU(3) CQSM,
the mean field approximation and the cranking method for the non-hedgehog soliton are introduced.
Section~\ref{sec:fluc} is the main part of this paper.
At first, we define the fluctuations around the soliton and obtain the effective action for the fluctuations.
Next the normalization and the overlap integral between the fluctuations are shown.
In Sec.~\ref{sec:num_results} we give the numerical results.
Section~\ref{sec:summary} is the summary of this paper.

\section{Casimir energy\label{sec:casimir_energy}}
The Casimir energy due to the soliton is given by a difference
between the two types of the zero point energy:
\begin{equation}
	\Delta E = \half \sum_{i} \omega_{i} - \half \sum_{j} \omega_{j}^{(0)},
\end{equation}
where $\omega_{i}$ are the energy eigenvalues of the fluctuations around the soliton
and $\omega_{j}^{(0)}$ denote ones in the absence of the soliton.
However, this expression diverges quadratically in $3+1$ dimensions.
Holzwarth has shown that the ultraviolet divergence requires the three subtraction terms~\cite{rf:Holzwarth94}.
After the subtractions, the finite expression is given by
\begin{equation}
	\Delta E = \half \sum_{i} \left\{
		\omega_{i}
		-\frac{1}{8} \sum_{j} \omega_{j}^{(0)}
		\left| \braketi{\tilde{z}(\omega_{i})}{\tilde{z}^{(0)}(\omega^{(0)}_{j})} \right|^{2}
		\left[ 
					3+6\left(\omega_{i}/\omega_{j}^{(0)}\right)^{2}
					-\left(\omega_{i}/\omega_{j}^{(0)}\right)^{4}
		\right]
	\right\},
\label{eq:casimir}
\end{equation}
where $\ket{\tilde{z}(\omega_{i})}$ and $\ket{\tilde{z}^{(0)}(\omega^{(0)}_{j})}$ are
the energy eigenstates of the fluctuations with the $\omega_{i}$ and $\omega^{(0)}_{j}$ respectively,
and $\braketi{\tilde{z}(\omega_{i})}{\tilde{z}^{(0)}(\omega^{(0)}_{j})}$
is the overlap integral between these states.
This is a $3+1$ dimensional generalization of the result in a $1+1$ case~\cite{rf:Cahill76}.

In this paper we study $\Delta E$ for the octet and decuplet baryons in the CQSM.
In the Skyrme model~\cite{rf:Holzwarth94} and the NJL model~\cite{rf:Weigel95},
$\Delta E$ is dominated by the contributions of the zero modes $\omega_{j} = 0$ and take a negative value
for the $N$ and the $\Delta$.
Thus we concentrate on the zero modes contributions:
\begin{equation}
	\Delta E \approx -\frac{3}{16} \sum_{\omega_{i}=0} \sum_{j}
		\omega_{j}^{(0)} \left| \braketi{\tilde{z}(\omega_{i})}{\tilde{z}^{(0)}(\omega^{(0)}_{j})} \right|^{2},
\label{eq:casimir_zeromode}
\end{equation}
where $\sum_{\omega_{i}=0}$ means that the mode sum for $\omega_{i}$ is restricted to the zero modes.

Equations~(\ref{eq:casimir}) and (\ref{eq:casimir_zeromode}) are finite objects
for the renormalizable theory in $3+1$ dimensions.
In this paper, however we study the CQSM which is a cutoff theory.
Thus the sums in these equations should be terminated at a physical cutoff point.
In addition the overlap integrals $\braketi{\tilde{z}(\omega_{i})}{\tilde{z}^{(0)}(\omega^{(0)}_{j})}$
should be defined by the quantities in this mode.
Therefore these equations are still formal definitions of the Casimir energy.
The cutoff procedure and the overlap integrals are defined in Sec.~\ref{sec:fluc}.

\section{SU(3) chiral quark soliton model\label{sec:su3_cqsm}}
The low energy baryonic states can be obtained through the study of a correlation function
for the quark operators~\cite{rf:Gupta87}:
\begin{eqnarray}
	J_{\Psi}(x) &=& \frac{1}{N_{c}!}
		\varepsilon_{\alpha_{N_{c}} \dots \alpha_{1}} \Gamma_{\Psi}^{f_{N_{c}} \dots f_{1}}
		\psi_{\alpha_{N_{c}}f_{N_{c}}}(x) \dots \psi_{\alpha_{1}f_{1}}(x),\\
	J_{\Psi}^{*}(y) &=& \frac{1}{N_{c}!}
		\varepsilon_{\beta_{N_{c}} \dots \beta_{1}} \Gamma_{\Psi}^{g_{N_{c}} \dots g_{1}*}
		\psi_{\beta_{1}f_{1}}^{*}(y) \dots \psi_{\beta_{N_{c}}f_{N_{c}}}^{*}(y),
\end{eqnarray}
where $N_{c}$ is the number of color, $\alpha$ and $\beta$ are color indices,
both $f$ and $g$ are the spin and flavor indices, and $\Gamma_{\Psi}$ is the symmetric tensor
representing the baryon state $\Psi$ in spin and flavor space.
In the CQSM the correlation function is given by
\begin{equation}
	\langle J_{\Psi}(x) J_{\Psi}^{*}(y) \rangle =
	\frac{1}{N_{c}!} \Gamma_{\Psi}^{f_{N_{c}} \dots f_{1}} \Gamma_{\Psi}^{g_{N_{c}} \dots g_{1}*}
	\int DU e^{i S_{F}\left[ U \right]} \prod^{N_{c}}_{j=1} \braketo{xf_{j}}{\frac{i}{i\partial_{t}-H}}{yg_{j}},
\label{eq:corr_func}
\end{equation}
where $U$ is the SU(3) chiral meson field and $S_{F}[U]$ is the effective action for $U$.
The quark Hamiltonian $H$ is given by
\begin{equation}
	H = \frac{1}{i} {\bm \alpha} \cdot \nabla + \beta \left(MU^{\gamma_{5}}+\hat{m}\right),
\end{equation}
where $U^{\gamma_{5}} = \frac{1+\gamma_{5}}{2} U + \frac{1-\gamma_{5}}{2} U^{\dagger}$,
$M$ is the dynamically generated quark mass, and $\hat{m}$ is the current quark mass matrix:
\begin{equation}
	\hat{m} = {\rm diag}(m_{u},m_{u},m_{s}) = m_{0} \lambda_{0} + m_{8} \lambda_{8}.
\label{eq:quark_mass}
\end{equation}
Then the effective action $S_{F}[U]$ is defined by the functional determinant for the quark fields,
\begin{equation}
	iS_{F}[U] = N_{c} \log \det (i\partial_{t}-H).
\end{equation}

We assume the so-called cranking form~\cite{rf:Adkins83,rf:Braaten88}
for the meson field:
\begin{equation}
	U^{\gamma_{5}}({\bf r},t) = {\cal A}(t){\cal B}^{\dagger}(t)
		U^{\gamma_{5}}_{0}({\bf r}) {\cal B}(t){\cal A}^{\dagger}(t),
\label{eq:aua}
\end{equation}
where $U^{\gamma_{5}}_{0}({\bf r})$ is the static meson field in the body fixed frame of the soliton,
${\cal A}(t)$ denotes the adiabatic rotation of the system in the SU(3) flavor space,
and ${\cal B}(t)$ describes the spatial rotation.
Then, the effective action for $U^{\gamma_{5}}$ is reduced to 
\begin{equation}
	i S_{F} = N_{c} \log\det \left(
				i\partial_{t}-H'-V_{A}
	\right),
\label{eq:action}
\end{equation}
where 
\begin{equation}
	V_{A} = -i {\cal A}^{\dagger}\dot{\cal A}-i {\cal B}\dot{{\cal B}}^{\dagger},
\end{equation}
and $H'$ is the rotated quark Hamiltonian with the meson fluctuations defined below.

For $U^{\gamma_{5}}_{0}$, we assume the embedding of the SU(2) field
to the SU(3) matrix but do not assume the hedgehog shape~\cite{rf:Akiyama03}:
\begin{equation}
	U^{\gamma_{5}}_{0}({\bf r}) = e^{i\gamma_{5} F({\bf r}) \hat{\Lambda}({\bf r})},
\label{eq:u_gamma_5}
\end{equation}
where $F$ is the radial component of the profile function, 
\begin{equation}
	\hat{\Lambda}({\bf r}) = \sum_{m} \hat{\Lambda}^{m}({\bf r}) \lambda_{m} \hspace{5mm} (m = \pm 1, 0),
\label{eq:lambda}
\end{equation}
$\hat{\Lambda}^{m}$ is the contravariant spherical component~\cite{rf:Varshalovich88}
of a unit vector in isospin space, and $\lambda_{m}$ is the covariant one
of the SU(2) subalgebra of the Gell-Mann matrices.
The transformations between the contravariant and the covariant spherical components
are given by
\begin{eqnarray}
	\Lambda_{m} &=& \sum_{m'} g_{mm'} \Lambda^{m'},\\
	\lambda^{m} &=& \sum_{m'} g^{mm'} \lambda_{m'},
\end{eqnarray}
where
\begin{equation}
	g_{mm'} = g^{mm'} = (-1)^{m} \delta_{m,-m'}
\end{equation}

We write the flavor rotation~\cite{rf:Kaplan90} as
\begin{equation}
	{\cal A}(t) = \left(\begin{array}{cc}
					A(t)		& 0\\
					0^{\dagger} & 1
			\end{array}\right) A_{s}(t),
\label{eq:A_param}
\end{equation}
where $A$ describes the rotation in SU(2) flavor space and 
$A_{s}$ represents the rotation into the strange directions.
Furthermore we parameterize $A_{s}(t)$ as
\begin{equation}
	A_{s}(t) = \exp i \left(
				\begin{array}{cc}
				0 & \sqrt{2} D(t)\\
				\sqrt{2} D^{\dagger}(t) & 0
				\end{array}
			\right),
\label{eq:As_param}
\end{equation}
where $D = (D_{1}, D_{2})^{T}$ is the isodoublet spinor.
In Ref.~\cite{rf:Westerberg94}, it is argued that $D \sim 1/\sqrt{N_{c}}$
in the large $N_{c}$ limit due to the Wess-Zumino term,
even if the strange quark mass is light.
We also employ the classification and treat $D$ perturbatively. 

From the current quark mass matrix (\ref{eq:quark_mass}),
we obtain the rotated one:
\begin{equation}
	\hat{m}' = {\cal A}^{\dagger} \hat{m} {\cal A}
	= m_{0} \lambda_{0} + m_{8} D^{(8)}_{8\mu}(A_{s}) \lambda_{\mu},
\label{eq:rot_mq}
\end{equation}
where $D^{(8)}_{\mu\nu}(A_{s})$ ($\mu,\nu = 1,2,\dots,8$) are the Wigner \textit{D}
functions of $A_{s}$ in the adjoint representation:
\begin{equation}
	D^{(8)}_{\mu\nu}(A_{s}) = \frac{1}{2} {\rm tr} \left(
			A_{s}^{\dagger} \lambda_{\mu} A_{s} \lambda_{\nu}
		\right).
\label{eq:d_func}
\end{equation}
The value of $m_{8}$ represents the strength of the flavor SU(3) symmetry breaking. 

Here, we define the following quantities:  
\begin{eqnarray}
	\kappa_{0} &\equiv& 2D^{\dagger} D,\\
	\kappa_{3} &\equiv& 2D^{\dagger} \tau_{3} D.
\end{eqnarray}
Suppose that the collective variables ${\cal A}$ and ${\cal B}$ are quantized
and $\ket{B}$ as an eigenstate of a collective Hamiltonian.
If $\ket{B}$ points to a specific direction in the isospin space,
the order parameters:
\begin{eqnarray}
 	\kappa_{B0} &=& \braketo{B}{\kappa_{0}}{B},
\label{eq:kappa0}\\
	\kappa_{B3} &=& \braketo{B}{\kappa_{3}}{B}
\label{eq:kappa3}
\end{eqnarray}
have nonzero values~\cite{rf:Akiyama03}. 
Then, the expectation value $\braketo{B}{\hat{m}'}{B}$ may be approximated by
\begin{equation}
	\hat{m}_{B} = m_{0} \lambda_{0}
		+ m_{8} \lim_{\kappa_{0,3} \rightarrow \kappa_{B0,3}}
		\left[ D^{(8)}_{83}(A_{s}) \lambda_{3} + D^{(8)}_{88}(A_{s}) \lambda_{8}\right].
\label{eq:mq_eff}
\end{equation}
The matrix $\hat{m}_{B}$ is diagonal and
its eigenvalues $(m_{Bu},m_{Bd},m_{Bs})$ are the effective quark masses in the body fixed frame.
The SU(2) quarks with masses ($m_{Bu}$,$m_{Bd}$) interact with the chiral field (\ref{eq:u_gamma_5}).
On the other hand, the quark with $m_{Bs}$ decouples in the classical soliton.

In Ref.~\cite{rf:Akiyama03,rf:Akiyama04},
we obtain that $\abs{\kappa_{B3}} \approx \kappa_{B0}$ for the octet and decuplet baryons
and the value of $\kappa_{B0}$ changes according to the strangeness.
For $\kappa_{B3} \approx -\kappa_{B0}$, the effective quark masses become
\begin{subequations}
\begin{eqnarray}
	m_{Bu} &=& m_{u},\\
	m_{Bd} &=& m_{u} \cos^{2} \sqrt{\kappa_{B0}} + m_{s} \sin^{2} \sqrt{\kappa_{B0}},\\
	m_{Bs} &=& m_{s} + \frac{1}{6} (m_{u}+m_{d}-2m_{s}) \sin^{2} \sqrt{\kappa_{B0}}.
\end{eqnarray}
\label{eq:mbq_eff1}
\end{subequations}
For $\kappa_{B3} \approx \kappa_{B0}$,
\begin{subequations}
\begin{eqnarray}
	m_{Bu} &=& m_{u} \cos^{2} \sqrt{\kappa_{B0}} + m_{s} \sin^{2} \sqrt{\kappa_{B0}},\\
	m_{Bd} &=& m_{u},\\
	m_{Bs} &=& m_{s} + \frac{1}{6} (m_{u}+m_{d}-2m_{s}) \sin^{2} \sqrt{\kappa_{B0}}.
\end{eqnarray}
\label{eq:mbq_eff2}
\end{subequations}

The important points in the present context are that 
one of the SU(2) quarks always has a light mass $m_{u}$ and
the other becomes heavy as $\kappa_{B0}$ grows.
In the CQSM, the pseudoscalar fields are auxiliary and
the corresponding mesons are bound states of the quarks in the body fixed frame.
Thus, in our treatment, the mesons would be not only the $\pi$ but also the $K$ and $\eta$
according to the strangeness of the soliton.

\section{Fluctuations\label{sec:fluc}}
Given $U_{0}^{\gamma_{5}}$, we evaluate the effects of the fluctuations of the pseudoscalar meson fields
by writing~\cite{rf:Weigel95}
\begin{equation}
	U_{0}^{\gamma_{5}}({\bf r}) \Rightarrow 
	U_{z}^{\gamma_{5}}({\bf r}, t) =
		e^{i \gamma_{5} F({\bf r}) \hat{\Lambda}({\bf r})/2} \cdot
		e^{i \gamma_{5} z({\bf r}, t)} \cdot
		e^{i \gamma_{5} F({\bf r}) \hat{\Lambda}({\bf r})/2},
\end{equation}
where $z({\bf r},t)$ are the small amplitude fluctuations and have three components in isospin space,
\begin{equation}
	z({\bf r}, t) = \sum_{m} z^{m}({\bf r}, t) \lambda_{m} \hspace{5mm} (m=\pm 1,0).
\label{eq:fluc_z}
\end{equation}
Here, we do not include the genuine SU(3) fluctuations $\sum_{a=4}^{7} z^{a} \lambda_{a}$ in $z({\bf r}, t)$,
since we have already incorporated the strangeness degrees of freedom by Eq.~(\ref{eq:As_param}).
The fluctuations $z({\bf r}, t)$ are induced in the body fixed frame of the soliton
and consist of the quarks with the masses $m_{Bu}$ and $m_{Bd}$.
Thus, the dispersion relations of the fluctuations also change according to the strangeness.

We expand the effective action up to quadratic order in power of $z({\bf r}, t)$.
Then the change of the chiral field due to the fluctuations is given by
\begin{equation}
	U_{z}^{\gamma_{5}}-U_{0}^{\gamma_{5}} =
		\sum_{m} \left[
			E_{m}\left({\bf r}\right) z^{m}({\bf r}, t)
			-\half P_{SU2} U_{0}^{\gamma_{5}}\left({\bf r}\right) z_{m}({\bf r}, t) z^{m}({\bf r}, t)
		\right],
\end{equation}
where $P_{SU2}$ is the projection operator $P_{SU2} = {\rm diag}\left(1,1,0\right)$ for SU(2) subspace
and
\begin{equation}
	E_{m}\left({\bf r}\right) =
		\hat{\Lambda}_{m} \left(-P_{SU2} \sin F + i \gamma_{5} \hat{\Lambda} \cos F\right)
		+ i \gamma_{5} \left(\lambda_{m}-\hat{\Lambda}_{m}\hat{\Lambda}\right).
\end{equation}

Using these quantities, the rotated quark Hamiltonian $H'$ with the meson fluctuations
in Eq.~(\ref{eq:action}) is given by
\begin{eqnarray}
	H' &=& H'_{0} + V_{S},
\label{eq:rot_hq}\\
	H'_{0} &=& \frac{1}{i} {\bm \alpha} \cdot \nabla + 
		\beta \left(MU^{\gamma_{5}}_{0}+\hat{m}_{B}\right),
\label{eq:rot_hq0}\\
	V_{S} &=& \beta \left(\hat{m}' - \hat{m}_{B}\right)
		+ \beta M \left(U_{z}^{\gamma_{5}}-U_{0}^{\gamma_{5}}\right),
\end{eqnarray}
where $(\hat{m}' - \hat{m}_{B})$ determines the fluctuation around the mean field $\hat{m}_{B}$,
and $H'_{0}$ contains the effects of the flavor SU(3) symmetry breaking
through $\hat{m}_{B}$. 

We expand the correlation function (\ref{eq:corr_func}) with respect to $V_{S}$ and $V_{A}$
around the eigenstates of the $H'_{0}$:
\begin{equation}
			H'_{0} \ket{a} = \epsilon_{a} \ket{a} 
\end{equation}
to obtain the effective action for the collective motions of the soliton
and the fluctuations around the soliton.
In this expansion we use the Schwinger proper time regularization~\cite{rf:Schwinger51,rf:Ebert86,rf:Diakonov88,rf:Weigel95}.
In this paper, we study only on the fluctuation dependent part of the action.
The other parts have been studied in Ref.~\cite{rf:Akiyama03,rf:Akiyama04} for non-hedgehog soliton.

The action up to quadratic order of the fluctuations
\begin{equation}
	\tilde{z}^{m}\left({\bf r},\omega\right) = \int dt e^{i \omega t} z^{m}({\bf r},t),
\end{equation}
becomes
\begin{eqnarray}
	S_{z} &=& \int \frac{d\omega}{2\pi} \int d^{3}r \int d^{3}r' \sum_{m,n}
		\half \tilde{z}^{m}\left({\bf r},+\omega\right) \tilde{z}^{n}\left({\bf r}',-\omega\right)
		\left[
			\Phi^{(2)}_{mn}\left({\bf r}, {\bf r}', \omega\right)+
			g_{mn} \delta^{(3)}\left({\bf r}-{\bf r}'\right) \Phi^{(1)}\left({\bf r}\right)
		\right] \nonumber\\
		&&-\int \frac{d\omega}{2\pi} \int d^{3}r \sum_{m}
			\tilde{z}^{m}\left({\bf r},\omega\right) J_{m}\left({\bf r},\omega\right).
\label{eq:sz_orig}
\end{eqnarray}
Here $\Phi^{(1)}\left({\bf r}\right)$ and $\Phi^{(2)}_{mn}\left({\bf r}, {\bf r}', \omega\right)$
are the local and bilocal kernels respectively:
\begin{eqnarray}
	\Phi^{(1)}\left({\bf r}\right) &=& M \sum_{a}
		\rho^{R}\left(\epsilon_{a}, \Lambda\right)
		\bar{\psi}_{a}\left({\bf r}\right)
			P_{SU2} U_{0}^{\gamma_{5}}\left({\bf r}\right)
		\psi_{a}\left({\bf r}\right),
\label{eq:localker}\\
	\Phi^{(2)}_{mn}\left({\bf r}, {\bf r}', \omega\right) &=& M^{2} \sum_{a,b}
		\rho_{\Gamma_{2}}\left(\epsilon_{a},\epsilon_{b},\omega,\Lambda\right)
		\bar{\psi}_{a}\left({\bf r}\right) E_{m}\left({\bf r}\right) \psi_{b}\left({\bf r}\right)
		\bar{\psi}_{b}\left({\bf r}'\right) E_{n}\left({\bf r}'\right) \psi_{a}\left({\bf r}'\right),
\label{eq:bilocalker}
\end{eqnarray}
where $\psi_{a}\left({\bf r}\right) = \braketi{{\bf r}}{a}$, $\Lambda$ is the cutoff parameter,
and $\rho^{R}$ and $\rho_{\Gamma_{2}}$ are the cutoff functions shown in Appendix~\ref{sec:cutoff}.
In addition, $J_{m}\left({\bf r}, \omega\right)$ are source functions for the fluctuations defined by
\begin{eqnarray}
	J_{m}\left({\bf r}, \omega\right) &=&
		-M \sum_{a} \rho^{R}\left(\epsilon_{a}, \Lambda\right)
			\bar{\psi}_{a}\left({\bf r}\right) E_{m}\left({\bf r}\right) \psi_{a}\left({\bf r}\right) \nonumber\\
		&&-M \tilde{\sigma}^{\mu}\left(-\omega\right) \sum_{a,b}
			\rho_{\Gamma_{2}}\left(\epsilon_{a},\epsilon_{b},\omega,\Lambda\right)
			\bar{\psi}_{a}\left({\bf r}\right) E_{m}\left({\bf r}\right) \psi_{b}\left({\bf r}\right)
			\braketo{b}{\beta T_{\mu}}{a},
\label{eq:source}
\end{eqnarray}
with
\begin{equation}
	\tilde{\sigma}^{\mu}\left(\omega\right) = \int dt e^{i \omega t} {\rm tr} \lambda_{\mu} \left[\hat{m}'(t)-\hat{m}_{B}\right].
\label{eq:sigma}
\end{equation}

The first term of $J_{m}\left({\bf r}, \omega\right)$ disappears
due to the equations of motion for the profile function~\cite{rf:Reinhardt88}
of the non-hedgehog soliton~\cite{rf:Akiyama03}:
\begin{eqnarray}
		S \sin F &=& ({\bf P} \cdot \hat{\bf \Lambda}) \cos F,\\
		P_{m} &=& ({\bf P} \cdot \hat{\bf \Lambda}) \hat{\Lambda}_{m},
\end{eqnarray}
where
\begin{eqnarray}
		S\left({\bf r}\right) &=& \sum_{a} \rho^{R}\left(\epsilon_{a}, \Lambda\right)
				\bar{\psi}_{a}\left({\bf r}\right) \psi_{a}\left({\bf r}\right),
\label{eq:scalar_density}\\
		P_{m}\left({\bf r}\right) &=& \sum_{b} \rho^{R}\left(\epsilon_{a}, \Lambda\right)
				\bar{\psi}_{a}\left({\bf r}\right) \lambda_{m} \psi_{a}\left({\bf r}\right).
\label{eq:pscalar_density}
\end{eqnarray}
In the second term of $J_{m}\left({\bf r}, \omega\right)$,
$\tilde{\sigma}^{\mu}$ denote the fluctuations of the mass matrix (\ref{eq:rot_mq})
around the mean field states $\ket{B_{0}}$ of the collective Hamiltonian~\cite{rf:Akiyama04}
and approximately satisfy 
\begin{equation}
	\braketo{B_{0}}{\tilde{\sigma}^{\mu}\left(-\omega\right)}{B_{0}} = 0.
\end{equation}
Thus we finally obtain
\begin{equation}
	\braketo{B_{0}}{J_{m}\left({\bf r},\omega\right)}{B_{0}} = 0.
\end{equation}

Up to the same order of approximation as the mean field,
the effective action for the fluctuations is given by
\begin{equation}
	S_{z} = \int \frac{d\omega}{2\pi} \int d^{3}r \int d^{3}r' \sum_{m,n}
		\half \tilde{z}^{m}\left({\bf r},+\omega\right) \tilde{z}^{n}\left({\bf r}',-\omega\right)
		\left[
			\Phi^{(2)}_{mn}\left({\bf r}, {\bf r}', \omega\right)+
			g_{mn} \delta^{(3)}\left({\bf r}-{\bf r}'\right) \Phi^{(1)}\left({\bf r}\right)
		\right].
\label{eq:sz_reduce}
\end{equation}

\subsection{Electric and Magnetic fluctuations}
The kernels $\Phi^{(1)}$ and $\Phi^{(2)}$ are composed of a large number of the quark states
and numerically calculated in our treatment.
In this process, the symmetries of the soliton are very helpful.
The non-hedgehog soliton in Ref.~\cite{rf:Akiyama03} is axial symmetric
and the profile function satisfies
\begin{eqnarray}
	F\left(r,\theta,\varphi\right) &=& F\left(r,\theta,0\right),
\label{eq:prof_f}\\
	\hat{\Lambda}_{m}\left(r,\theta,\varphi\right) &=&  \hat{\Lambda}_{m}\left(r,\theta,0\right) e^{i m \varphi}.
\label{eq:prof_t}
\end{eqnarray}
Since the third component of the grand spin ${\bf K}^{(q)} = {\bf J}^{(q)} + {\bf I}^{(q)}$
commutes with $H'_{0}$ [Eq.~(\ref{eq:rot_hq0})],
we can select the eigenstate $\ket{a}$ of $H'_{0}$ as the eigenstate of $K^{(q)}_{3}$,
\begin{equation}
	K^{(q)}_{3} \ket{a} = M_{a} \ket{a}.
\label{eq:k3_symmetry}
\end{equation}
In this case, the isoscalar-scalar $S^{(a,b)}$ and the isovector-pseudoscalar $P^{(a,b)}_{m}$
densities satisfy
\begin{eqnarray}
	S^{(a,b)}\left(r,\theta,\varphi\right) 
		&\equiv& \bar{\psi}_{a}\left({\bf r}\right) P_{SU2} \psi_{b}\left({\bf r}\right),\nonumber\\
		&=& e^{i \left(M_{b}-M_{a}\right) \varphi} S^{(a,b)}\left(r,\theta,0\right),
\label{eq:scalar}\\
	P^{(a,b)}_{m}\left(r,\theta,\varphi\right) 
		&\equiv& \bar{\psi}_{a}\left({\bf r}\right) i \gamma_{5} \lambda_{m} \psi_{b}\left({\bf r}\right),\nonumber\\
		&=& e^{i \left(M_{b}+m-M_{a}\right) \varphi} P^{(a,b)}_{m}\left(r,\theta,0\right).
\label{eq:pscalar}
\end{eqnarray}

From Eqs.~(\ref{eq:prof_f})-(\ref{eq:pscalar}),
the local kernel $\Phi^{(1)}$ is independent of the azimuth angle $\varphi$,
\begin{equation}
	\Phi^{(1)}\left(r,\theta,\varphi\right) = \Phi^{(1)}\left(r,\theta,0\right),
\end{equation}
and is accurately calculated for the baryon number $B = 1$ case. 
The part $(\bar{\psi}_{a} E_{m} \psi_{b})$ of $\Phi^{(2)}$ has a simple dependence on $\varphi$:
\begin{equation}
	\bar{\psi}_{a}\left(r,\theta,\varphi\right) E_{m}\left(r,\theta,\varphi\right) \psi_{b}\left(r,\theta,\varphi\right) =
		e^{i \left(M_{b}+m-M_{a}\right) \varphi}
		\bar{\psi}_{a}\left(r,\theta,0\right) E_{m}\left(r,\theta,0\right) \psi_{b}\left(r,\theta,0\right).
\end{equation}
However it is difficult to calculate $\Phi^{(2)}$ itself even numerically,
since the bilocal kernel $\Phi^{(2)}$ can contain any pairs of ($M_{a}$, $M_{b}$).
Thus, one adopts an ans\"{a}tz for the fluctuations $\tilde{z}$ and simplifies the action~(\ref{eq:sz_reduce}).

In the case of the hedgehog configuration, the fluctuations $\tilde{z}$ are expanded in $\hat{x}_{m} = x_{m}/r$
to separate the radial and angular dependences.
According to the parity, the expansion branches into two modes :
electric $\tilde{z}^{E}$ and magnetic $\tilde{z}^{M}$ fluctuations.
Then one usually adopt the following ans\"{a}tz~\cite{rf:Walliser84,rf:Holzwarth94,rf:Weigel95},
\begin{subequations}
\begin{eqnarray}
	\tilde{z}^{E}_{m}\left({\bf r},\omega\right) &=& \sum_{m_{1}} \left[
		\hat{x}_{m} \hat{x}_{m_{1}} \zeta^{Lm_{1}}(r,\omega)
		+\left(g_{mm_{1}}-\hat{x}_{m} \hat{x}_{m_{1}}\right) \zeta^{Tm_{1}}(r,\omega)
	\right],
\label{eq:ze_h}\\
	\tilde{z}^{M}_{m}\left({\bf r},\omega\right) &=& \sum_{m_{1},m_{2}}
		\varepsilon_{mm_{1}m_{2}} \hat{x}^{m_{1}} \zeta^{Mm_{2}}(r,\omega).
\label{eq:zm_h}
\end{eqnarray}
\label{eq:z_h}
\end{subequations}
Here, we employ the slightly different parametrization of $\tilde{z}^{E}_{m}$ from Ref.~\cite{rf:Weigel95},
because of the algebraic manageability.

For the non-hedgehog one, we take the straightforward generalization of Eq.~(\ref{eq:z_h}),
\begin{subequations}
\begin{eqnarray}
	\tilde{z}^{E}_{m}\left({\bf r},\omega\right) &=& \sum_{m_{1}} \left[
		\hat{\Lambda}_{m} \hat{\Lambda}_{m_{1}} \zeta^{Lm_{1}}(r,\theta,\omega)
		+\left(g_{mm_{1}}-\hat{\Lambda}_{m} \hat{\Lambda}_{m_{1}}\right) \zeta^{Tm_{1}}(r,\theta,\omega)
	\right],
\label{eq:ze_nh}\\
	\tilde{z}^{M}_{m}\left({\bf r},\omega\right) &=& \sum_{m_{1},m_{2}}
		\varepsilon_{mm_{1}m_{2}} \hat{\Lambda}^{m_{1}} \zeta^{Mm_{2}}(r,\theta,\omega),
\label{eq:zm_nh}
\end{eqnarray}
\label{eq:z_nh}
\end{subequations}
where the amplitudes $\zeta^{L,T,M}_{m}$ depend on $r$ and $\theta$.
This ans\"{a}tz is not so toy as the appearance.
Expansion of the fluctuations in terms of the vector spherical harmonics is given by
\begin{equation}
	\tilde{z}_{m}\left({\bf r},\omega\right) =
		\sum_{J,J_{3}} \sum_{\lambda=-1}^{1} \left[{\bf Y}^{(\lambda)}_{JJ_{3}}\left(\theta,\varphi\right)\right]_{m}
		\zeta^{(\lambda)}_{JJ_{3}}(r,\omega),
\label{eq:z_harmonic_expand}
\end{equation}
where $\zeta^{(\lambda)}_{JJ_{3}}(r,\omega)$ is a function of $r$.
By exchanging the order of the summation, we obtain
\begin{equation}
	\tilde{z}_{m}\left({\bf r},\omega\right) =
		\sum_{J_{3}} e^{i\left(J_{3}+m\right) \varphi} \zeta_{J_{3},m}(r,\theta,\omega),
\label{eq:z_j3_expand}
\end{equation}
where
\begin{equation}
	\zeta_{J_{3},m}(r,\theta,\omega) =
		\sum_{J \geq \abs{J_{3}},\lambda} \left[{\bf Y}^{(\lambda)}_{JJ_{3}}\left(\theta,0\right)\right]_{m}
		\zeta^{(\lambda)}_{JJ_{3}}(r,\omega).
\label{eq:z_j3_expand_cof}
\end{equation}
Because of Eq.~(\ref{eq:prof_t}), Eq.~(\ref{eq:z_nh}) corresponds to Eq.~(\ref{eq:z_j3_expand})
with $\abs{J_{3}} \leq 1$.
Then the $\theta$ dependences of $\zeta^{L,T,M}_{m}$ correspond to the high $J$ components
in Eq.~(\ref{eq:z_j3_expand_cof}). And we can incorporate the complicated variations
in the $\theta$ direction due to the deformation of the soliton.
Thus, Eq.~(\ref{eq:z_nh}) would work well for low energy excitations,
especially the zero modes.

Using Eq.~(\ref{eq:ze_nh}) in Eq.~(\ref{eq:sz_reduce}) and integrating over $\varphi$,
we obtain the effective action for the electric fluctuations: 
\begin{eqnarray}
	S_{z}^{E} &=& \half \int \frac{d\omega}{2\pi} \sum_{m}
		\left[
			\left(2\pi\right)^{2} \int r^{2} \sin \theta dr d\theta
			\int r'^{2} \sin \theta' dr' d\theta' \right. \nonumber\\
			&&\zeta^{E}_{m}\left(r,\theta,\omega\right)^{T}
				\Phi^{(2)E}_{m}\left(r,\theta,r',\theta',\omega\right)
				\zeta^{Em}\left(r',\theta',-\omega\right) \nonumber\\
			&&\left.
			+2\pi \int r^{2} \sin \theta dr d\theta
				\zeta^{E}_{m}\left(r,\theta,\omega\right)^{T}
				\Phi^{(1)E}_{m}\left(r,\theta\right)
				\zeta^{Em}\left(r,\theta,-\omega\right)
		\right],
\label{eq:sz_e}
\end{eqnarray}
where 
\begin{equation}
	\zeta^{Em}\left(r,\theta,\omega\right) =
		\left(
			\begin{array}{c}
			\zeta^{Lm}\left(r,\theta,\omega\right)\\
			\zeta^{Tm}\left(r,\theta,\omega\right)				
			\end{array}
		\right),
\end{equation}
and $\Phi^{(1)E}$ and $\Phi^{(2)E}$ are the $2\times2$ electric local and bilocal kernels
given in Appendix~\ref{sec:bs_kernel} respectively. 
Using Eq.~(\ref{eq:zm_nh}), the effective action for the magnetic fluctuations becomes
\begin{eqnarray}
	S_{z}^{M} &=& \half \int \frac{d\omega}{2\pi} \sum_{m}
		\left[
			\left(2\pi\right)^{2} \int r^{2} \sin \theta dr d\theta
			\int r'^{2} \sin \theta' dr' d\theta' \right. \nonumber\\
			&&\zeta^{M}_{m}\left(r,\theta,\omega\right)
				\Phi^{(2)M}_{m}\left(r,\theta,r',\theta',\omega\right)
				\zeta^{Mm}\left(r',\theta',-\omega\right)\nonumber\\
			&&\left.
			+2\pi \int r^{2} \sin \theta dr d\theta
				\zeta^{M}_{m}\left(r,\theta,\omega\right)
				\Phi^{(1)M}_{m}\left(r,\theta\right)
				\zeta^{Mm}\left(r,\theta,-\omega\right)
		\right],
\label{eq:sz_m}
\end{eqnarray}
where $\Phi^{(1)M}$ and $\Phi^{(2)M}$ are the magnetic local and bilocal kernels
given in Appendix~\ref{sec:bs_kernel} respectively.
Here the amplitudes $\zeta^{E,M}_{m}$ with the different spherical indices decouple each other.

The equations of motion for fluctuations (Bethe-Salpeter equations) are obtained from
the minimization of $S_{z}^{E}$ and $S_{z}^{M}$ with respect to $\zeta^{E}_{m}$ and $\zeta^{M}_{m}$:
\begin{subequations}
\begin{eqnarray}
		2\pi \int r'^{2} \sin \theta' dr' d\theta'
		\Phi^{(2)E}_{m}\left(r,\theta,r',\theta',\omega\right)
		\zeta^{Em}\left(r',\theta',-\omega\right)\nonumber\\
		+\Phi^{(1)E}_{m}\left(r,\theta\right)
		\zeta^{Em}\left(r,\theta,-\omega\right) = 0,\\
		2\pi \int r'^{2} \sin \theta' dr' d\theta'
		\Phi^{(2)M}_{m}\left(r,\theta,r',\theta',\omega\right)
		\zeta^{Mm}\left(r',\theta',-\omega\right)\nonumber\\
		+\Phi^{(1)M}_{m}\left(r,\theta\right)
		\zeta^{Mm}\left(r,\theta,-\omega\right) = 0.
\end{eqnarray}
\label{eq:bs_equation}
\end{subequations}
In addition, the boundary conditions
\begin{subequations}
\begin{eqnarray}
	\zeta^{Em}\left(R,\theta,-\omega\right) &=& 0,\\
	\zeta^{Mm}\left(R,\theta,-\omega\right) &=& 0,
\end{eqnarray}
\label{eq:boundary_cond}
\end{subequations}
are required at the surface ($r = R$) of a sufficient large spherical box,
in which the numerical calculation is performed.
These equations determine the fluctuation amplitudes $\zeta^{E,M}_{m}$ and its eigenvalues $\omega$.

As referred in Sec.~\ref{sec:casimir_energy}, we restrict consideration to
only zero modes ($\omega_{j} = 0$) in $B = 1$ sector.
Since
\begin{eqnarray}
	\Phi^{(2)E}_{m}\left(r,\theta,r',\theta',0\right) &=& \Phi^{(2)E}_{m}\left(r',\theta',r,\theta,0\right),\\
	\Phi^{(2)M}_{m}\left(r,\theta,r',\theta',0\right) &=& \Phi^{(2)M}_{m}\left(r',\theta',r,\theta,0\right),
\end{eqnarray}
we can select the phases of the amplitudes so that
\begin{eqnarray}
	\zeta^{E}_{m}(r,\theta,0) &=& \zeta^{E}_{-m}(r,\theta,0),\\
	\zeta^{M}_{m}(r,\theta,0) &=& \zeta^{M}_{-m}(r,\theta,0).
\end{eqnarray}

\subsection{Normalization}
The Casimir energy (\ref{eq:casimir_zeromode}) needs determining of the normalizations
and the overlap integrals between the fluctuations in the $B = 0, 1$ sectors.
Weigel \textit{et al.} define them with the Bethe-Salpeter kernels in the NJL model~\cite{rf:Weigel95}.
Although their calculation is restricted to the case of the hedgehog soliton,
the procedure can be extended to the non-hedgehog one.
For the completeness, we repeat their procedure here.
At first, we define the metric tensor for each spherical index $m$
of the electric and magnetic fluctuations,
\begin{equation}
	{\cal M}^{ab}_{m}\left({\bf r},{\bf r}',\omega_{j}\right) =
		\frac{\partial}{\partial \omega^{2}}
		\left. \Phi^{(2)ab}_{m}\left({\bf r},{\bf r}',\omega\right) \right|_{\omega=\omega_{j}},
\end{equation}
where $\Phi^{(2)}_{m}$ is the $\Phi^{(2)E}_{m}$ or $\Phi^{(2)M}_{m}$.
The indices $a$,$b$ are only for the electric fluctuations
and refer to the element of $2\times2$ matrix,
and $\omega_{j}$ are eigenvalues of the fluctuations.
Then one can demand the normalization condition for the fluctuation amplitudes $\zeta^{E,M}_{m}$,
\begin{equation}
	\int d^{3}r d^{3}r'
		\sum_{a,b} \zeta^{a}_{m}({\bf r},\omega_{j})
		{\cal M}^{ab}_{m}\left({\bf r},{\bf r}',\omega_{j}\right)
		\zeta^{bm}({\bf r}',-\omega_{j}) =1.
\label{eq:norm_cond}
\end{equation}
The wave function of the fluctuations are given by
\begin{equation}
	\phi^{a,m}\left({\bf r},\omega_{j}\right) = \int d^{3}x \sum_{c}
		\left(\sqrt{{\cal M}_{m}}\right)^{ac}\left({\bf r},{\bf x},\omega_{j}\right)
		\zeta^{c,m}\left({\bf x},\omega_{j}\right),
\label{eq:wave_func}
\end{equation}
where $\sqrt{{\cal M}_{m}}$ is the square root of the metric tensor determined by
\begin{equation}
	{\cal M}^{ab}_{m}\left({\bf r},{\bf r}',\omega_{j}\right) = \int d^{3}x \sum_{c}
		\left(\sqrt{{\cal M}_{m}}\right)^{ac}\left({\bf r},{\bf x},\omega_{j}\right)
		\left(\sqrt{{\cal M}_{m}}\right)^{bc}\left({\bf r}',{\bf x},\omega_{j}\right).
\end{equation}
The normalization condition becomes
\begin{equation}
	\int d^{3} r \sum_{a} \phi^{a}_{m}({\bf r},\omega_{j}) \phi^{a,m}({\bf r},\omega_{j}) = 1.
\end{equation}
Finally the overlap integral  between the fluctuations in the $B = 0, 1$ sectors is defined by
\begin{equation}
	\braketi{\tilde{z}_{m}(\omega_{j})}{\tilde{z}^{(0)}_{m}(\omega^{(0)}_{j})} =
		\int d^{3}r \sum_{a} \phi^{a}_{m}({\bf r},\omega_{j}) \phi^{(0)a,m}({\bf r},\omega^{(0)}_{j}),
\end{equation}
where the quantities with the index $(0)$ is in $B = 0$ sector.

\subsection{Baryon number $B = 0$ sector\label{sec:b0_sector}}
In $B = 0$ sector, it is difficult to calculate accurately the kernels $\Phi^{(2)E,M}$,$\Phi^{(1)E,M}$
with the spherical plane wave basis~\cite{rf:Kahana84} in the large spherical box.
The difficulty is remarkable at the vicinity of the surface of the box and not negligible
since the fluctuations in $B = 0$ sector should freely propagate.

Instead, we evaluate the dispersion relations of the fluctuations with the help of the linear momentum basis.
This is systematically done by evaluating the action (\ref{eq:sz_reduce})
in the 4-dimensional momentum $k$ space,
\begin{eqnarray}
	S_{z} &=& \frac{N_{c}}{2} \int \frac{d^{4} k}{(4\pi)^{4}} \sum_{m} \tilde{z}_{m}(k) \tilde{z}^{m}(-k)
	\left[
		\frac{M}{4\pi^{2}} \sum_{q} m_{Bq} M_{Bq}^{2} \Gamma\left(-1,(M_{Bq}/\Lambda)^{2}\right)
	\right.\nonumber\\
	&&-\delta_{m0} \frac{M^{2}}{8\pi^{2}} k^{2} \int^{1}_{0} d\alpha
			\sum_{q} \Gamma\left(0,[M_{Bq}^{2}-\alpha (1-\alpha) k^{2}]/\Lambda^{2}\right) \nonumber\\
	&&\left.-\delta_{\abs{m}1} \frac{M^{2}}{4\pi^{2}} [k^{2}-(m_{Bu}-m_{Bd})^{2}]
		\int^{1}_{0} d\alpha
			\Gamma\left(0,[(1-\alpha) M_{Bu}^{2} + \alpha M_{Bd}^{2} - \alpha (1-\alpha) k^{2}]/\Lambda^{2}\right)
	\right],
\label{eq:sz_b0}
\end{eqnarray}
where $q = u,d$, and $M_{Bq}$ are the constituent quark masses in the body fixed frame given by
\begin{equation}
	M_{Bq} = M + m_{Bq},
\end{equation}
$m_{Bq}$ are the eigenvalues of the quark mass matrix (\ref{eq:mq_eff}),
and $\Gamma$ is the incomplete gamma function.

By solving the Bethe-Salpeter equation:
\begin{equation}
	\frac{\delta S_{z}}{\delta \tilde{z}_{m}(k)} = 0,
\label{eq:bs_eq_b0}
\end{equation}
we obtain the following dispersion relation,
\begin{equation}
	k^{2} = (\omega^{(0)})^{2}-\abs{\bf k}^{2} = W_{m}^{2},
\end{equation}
where $W_{m}$ is the mass of the fluctuation $\tilde{z}_{m}(k)$, $\omega^{(0)}$ is the energy eigenvalue,
and ${\bf k}$ is a 3-dimensional momentum.
The mass $W_{m}$ is the lower bound of the $\omega^{(0)}$ in this sector.
The Kronecker deltas $\delta_{m0}$ and $\delta_{\abs{m}1}$ in Eq.~(\ref{eq:sz_b0}) indicate
that the masses of the fluctuations take different values according to the spherical index $m$.
An important thing to be emphasized in the present context is that the mass $W_{m}$ varies
through $m_{Bq}$ according to the strangeness of the soliton.

On the other hand, the fluctuations are the composite particles and may decay into quark-antiquark pairs
at a threshold energy $\omega_{th}$, because of the lack of the confinement in this model.
If we obey the argument in Ref.~\cite{rf:Weigel95} as it is and define the threshold
so that the Feynman parameter ($\alpha$) integrals in Eq.~(\ref{eq:sz_b0}) diverge at $k^{2} = \omega_{th}^{2}$,
we will obtain $\omega_{th} = 2 \min(M_{Bu},M_{Bd})$ for $m = 0$ and $\omega_{th} = M_{Bu}+M_{Bd}$ for $m = \pm 1$.
In our approach, however, there are situations $M_{Bu} \gg M_{Bd}$ or $M_{Bu} \ll M_{Bd}$
for the hyperons (Sec.~\ref{sec:su3_cqsm}).
Then the "threshold" $\omega_{th} = M_{Bu}+M_{Bd}$ for $m = \pm 1$ exceeds $2 \min(M_{Bu},M_{Bd})$ largely,
at which the Dirac sea of the lighter quark may become unstable.
Thus the argument in Ref.~\cite{rf:Weigel95} is applicable only for the case $M_{Bu} \approx M_{Bd}$.
Instead, we employ only one threshold $\omega_{th} = 2 \min(M_{Bu},M_{Bd})$ for $m = 0, \pm 1$.
Since $\min(m_{Bu},m_{Bd}) \approx m_{u}$ for the octet and decuplet baryons
[Eqs.~(\ref{eq:mbq_eff1}),(\ref{eq:mbq_eff2})],
the fluctuation $\tilde{z}_{m}(\omega^{(0)})$ can be induced within the range:
\begin{equation}
	W_{m} \leq \omega^{(0)} \leq \omega_{th} = 2 (M+m_{u}).
\label{eq:energy_region}
\end{equation}

For the wave functions of the fluctuations, we employ the Klein-Gordon operator~\cite{rf:Weigel95}
as the (local+bilocal) kernel in Eq.~(\ref{eq:sz_reduce}),
\begin{equation}
	\Phi^{(2)}_{mn}({\bf r},{\bf r}',\omega)+g_{mn} \delta^{(3)}({\bf r}-{\bf r}') \Phi^{(1)}({\bf r})
		= g_{mn} \delta^{(3)}({\bf r}-{\bf r}') (\omega^{2}+\nabla^{2} - W_{m}^{2}),
\end{equation}
with the ans\"{a}tz (\ref{eq:z_h}), and the boundary condition (\ref{eq:boundary_cond}).
Explicit forms of the wave functions for $B = 0$ sector are given in Appendix~\ref{sec:b0_wf}.
The $\omega_{j}^{(0)}$ sum in Eq.~(\ref{eq:casimir_zeromode}) is performed
by the angular momentum (S,P,D-waves) and the momentum ($k = \abs{\bf k}$) sums.
Both S,D-waves are the electric fluctuations and P-wave is the magnetic one.
The momentum $k$ is discretized by the boundary condition (\ref{eq:boundary_cond}).
The energy region~(\ref{eq:energy_region}) defines the cutoff procedure
for the zero mode contributions to the Casimir energy.

\section{Numerical results\label{sec:num_results}}
We use the following numerical values for the input parameters:
the dynamically generated quark mass $M = 400$ MeV,
the current (u,s) quark masses $(m_{u}, m_{s}) = (15, 210)$ MeV,
and the cutoff parameter $\Lambda = 700$ MeV.
These parameters are adjusted to reproduce the empirical values
of the pion mass and the octet and decuplet baryon masses.

The order parameters $\kappa_{B0}$ and $\kappa_{B3}$ [Eqs.~(\ref{eq:kappa0}) and (\ref{eq:kappa3})]
are self-consistently determined in the mean field approximation~\cite{rf:Akiyama04}
for the octet and decuplet baryons and given in Table~\ref{tb:order_param}.
\begin{table}
	\caption{The order parameters $\kappa_{B0} = |\kappa_{B3}|$.
	$S$ is the strangeness of soliton.
	}
	\begin{ruledtabular}
	\begin{tabular}{cc}
	$S$	& $\kappa_{B0} = |\kappa_{B3}|$\\
	\hline
	 0	& 0.00\\
	-1	& 0.80\\
	-2	& 1.65\\
	-3	& 2.57
	\end{tabular}
	\end{ruledtabular}
\label{tb:order_param}
\end{table}

\begin{figure*}
\includegraphics{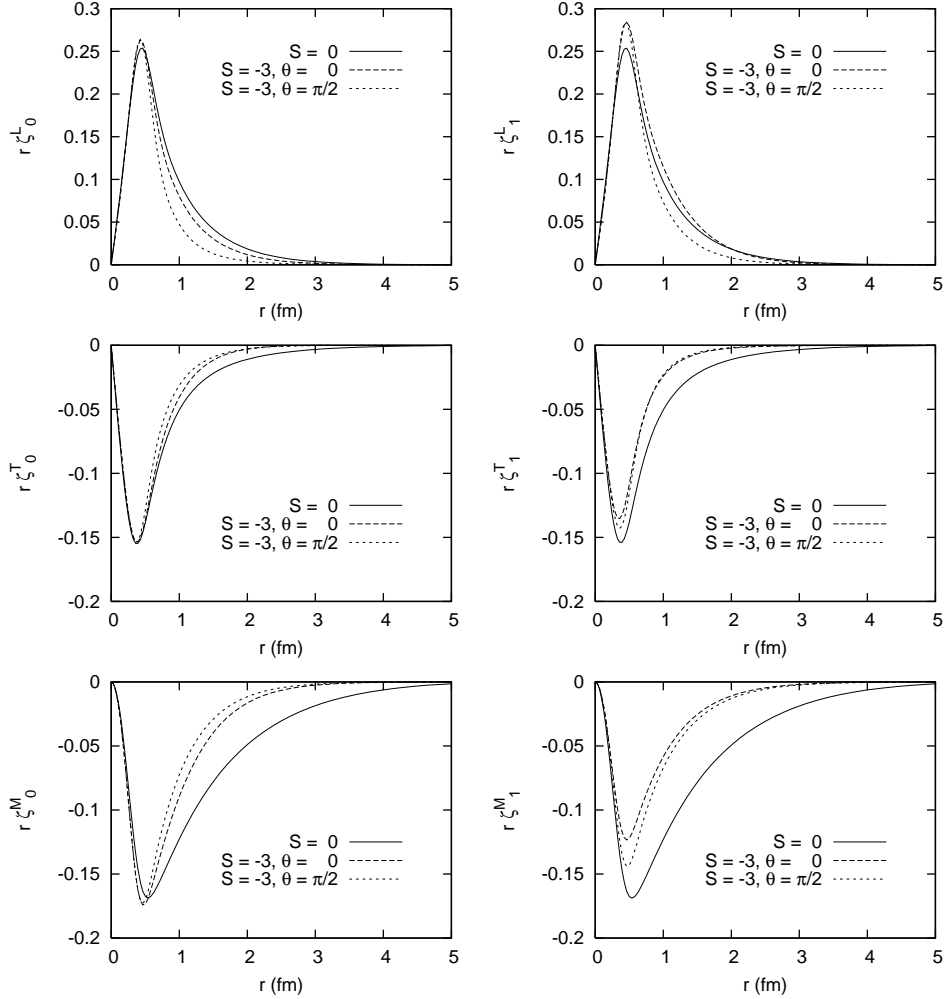}
	\caption{
		(Electric and magnetic fluctuation amplitudes)$\times r$
		 with $m = 0, +1$ in $B = 1$ sector :
		$r \zeta^{L}_{m}, r \zeta^{T}_{m}$, and $r \zeta^{M}_{m}$.
		The solid curves represent the $S = 0$ cases, 
		the dashed ones for $S = -3$ and $\theta = 0$, and
		the dotted ones for $S = -3$ and $\theta = \pi/2$.
	}
\label{fig:fluc_amp_s3}
\end{figure*}

We first show the amplitudes $\zeta^{L,T,M}_{m}$ of the zero mode fluctuations in the $B = 1$ sector.
These amplitudes are the solutions for $\omega = 0$ of the Bethe-Salpeter (BS) equations~(\ref{eq:bs_equation})
and are the boundary conditions~(\ref{eq:boundary_cond}),
and normalized by Eq.~(\ref{eq:norm_cond}).
Because of the  numerical reason, we calculate the fluctuation amplitudes multiplied by $r$.
It is enough to evaluate the Casimir energy~(\ref{eq:casimir_zeromode}). 
In Fig.~\ref{fig:fluc_amp_s3} the electric $r \zeta^{E}_{m} = (r \zeta^{L}_{m}, r \zeta^{T}_{m})$
and magnetic $r \zeta^{M}_{m}$ amplitudes with the spherical indices $m = 0, +1$
are displayed for $S = 0,-3$ cases.
The amplitudes for $S = -1, -2$ have a similar $r$ dependence to one for $S = -3$.
The $\theta$ dependences for $S = -1, -2$ are somewhat weaker than one for $S = -3$.

The amplitudes for $S = 0$ do not depend on $\theta$.
It is consistent with the hedgehog shape of the soliton.
Since the BS equations and the boundary conditions depend on $\theta$,
we can check the accuracy of the numerical calculation by the spherical symmetry.
The curves for $S \not= 0$ depend not only on $r$ but also $\theta$.
The $\theta$ dependence is not direct consequence from the non-hedgehog soliton,
because we expand the fluctuations $\tilde{z}$ in power of $\hat{\Lambda}$.
However it shows the self-consistency of the ans\"{a}tz~(\ref{eq:z_nh}).

In our treatment, the meson fluctuations are composed
of the quark and antiquark in the body fixed frame.
Since the quarks for $S \not= 0$ are heavier than ones for $S = 0$,
the fluctuations for $S \not= 0$ shrink toward the center of the soliton.
The same feature applies to the profile functions~\cite{rf:Akiyama03}.
The amplitudes, especially $\zeta^{M}_{m}$, indicate that the meson fluctuations for $S \not= 0$
are quite different from ones for $S = 0$.
In the momentum representation,  the distributions of the fluctuations for $S \not= 0$
shift to the high energy region comparing with the $S = 0$ case.
This nature of the amplitudes is important for the Casimir energy as noted below.

In the $B = 0$ sector, the energy eigenvalues $\omega^{(0)}$ of the fluctuations start from the mass $W_{m}$
and end at the threshold $\omega_{th} = 2(M+m_{u}) = 830$ MeV in Eq.~(\ref{eq:energy_region}).
The lower bounds $W_{m}$ depend on the spherical index $m$ and the strangeness of the soliton.
The values of $W_{m}$ are shown in Table~\ref{tb:fluc_mass}.
The u quark mass $m_{u}$ is chosen to fit $W_{m}$ with the pion mass for $S = 0$.
It is apparent in this sector that the masses of the fluctuations for $S \not= 0$ are heavier than for $S = 0$.
The masses $W_{+1}$ and $W_{-1}$ with the same strangeness are degenerate.
It is due to the charge (particle-antiparticle) symmetry of the action~(\ref{eq:sz_b0}). 
\begin{table}
	\caption{The masses $W_{m}$ of the meson fluctuations with spherical index $m$.
	$S$ is the strangeness of soliton.
	}
	\begin{ruledtabular}
	\begin{tabular}{ccc}
	$S$	& $W_{0}$ (MeV)	&$W_{\pm 1}$ (MeV)\\
	\hline
	 0	& 139 	& 139\\
	-1	& 284	& 311\\
	-2	& 315	& 369\\
	-3	& 320	& 383\\
	\end{tabular}
	\end{ruledtabular}
\label{tb:fluc_mass}
\end{table}

\begin{figure}
\includegraphics{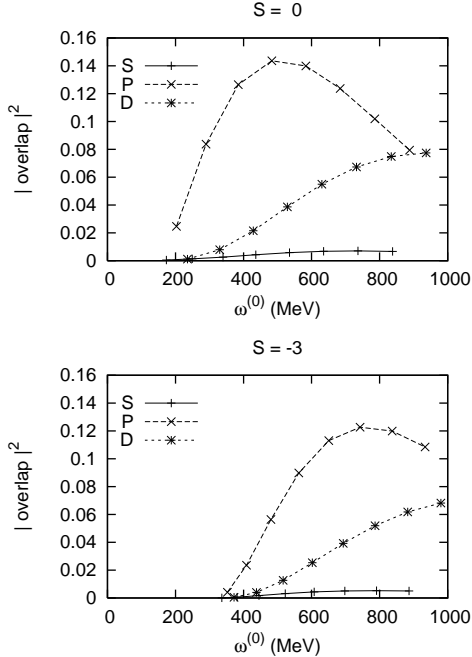}
	\caption{The $\omega^{(0)}$ dependence of the square of the overlap integral
	$|\braketi{\tilde{z}_{m}(\omega=0)}{\tilde{z}^{(0)}_{m}(\omega^{(0)})}|^{2}$
	with $m = 0$ for the strangeness $S = 0, -3$.
	The solid line is the S-wave contribution, the dashed line for P-wave,
	and the dotted line for D-wave.
	The radius of the large spherical box is chosen as $6$ fm.
	}
\label{fig:overlap}
\end{figure}

We are now ready to calculate the overlap integrals
$\braketi{\tilde{z}_{m}(\omega=0)}{\tilde{z}^{(0)}_{m}(\omega^{(0)})}$
between the zero mode fluctuations in $B = 1$ sector and the S,P,D-waves fluctuations in $B = 0$ sector.
These are calculated for individual spherical indices $m$.
Then the radius of the large spherical box is chosen as $R = 6$ fm.
Although the overlap integrals and the momenta given by Eq.~(\ref{eq:boundary_cond}) depend on the radius,
the Casimir energy (\ref{eq:casimir_zeromode}) do not in principle.
In Fig.~\ref{fig:overlap} we present the square of the overlap integrals as a function of $\omega^{(0)}$.
It is apparent that the D-wave contribution is dominating for electric fluctuations.
This result has already been reported for the non-strange baryons in Refs.~\cite{rf:Holzwarth94,rf:Weigel95}.

The sum over $\omega^{(0)}$ in Eq.~(\ref{eq:casimir_zeromode}) is restricted for $S \not= 0$
by comparison with $S = 0$.
It is because that the overlap integrals for $S \not= 0$ are shifted to the higher energy region
by the heavy masses, but the threshold $\omega_{th}$ is in common to the strangeness.

The Casimir energies $\Delta E$ for $S = (0, -1, -2, -3)$
are summarized in Table~\ref{tb:casimir_energy}.
$\abs{\Delta E}$ for $S \not= 0$ are about 80 MeV smaller than one for $S = 0$.
Thus the heavy fluctuations for $S \not= 0$ are suppressed.
It is likely from the physical point of view.
However these values are affected by the threshold $\omega_{th}$,
since the overlap integrals do not sufficiently fall off
in the vicinity of the $\omega^{(0)} = \omega_{th}$.
In Ref.~\cite{rf:Weigel95} this phenomenon is found for the electric modes.
In our approach, we encounter another origin: the heavy fluctuations for the hyperons.
For example, if we employ $\omega_{th} = M_{Bu} + M_{Bs}$
for the fluctuations with the spherical indices $m = \pm 1$,
the Casimir energies for the hyperons become comparable with ones for the non-strange baryons.
Although such a threshold is rejected from the physical argument in the Sec.~\ref{sec:b0_sector},
the overlap integrals should exclude the high energy modes normally.
However it might require the detailed knowledge on the confinement
and take us much beyond the scope of the present model.
Thus if one wants to extract the consistent physical informations from the present results,
the cutoff procedure for the fluctuations should be regarded as a part of the model.  
\begin{table}
	\caption{The Casimir energy for $B = 1$ sector.
	$S$ is the strangeness of soliton.
	$\Delta E_{e}$ is the contribution of the electric fluctuation.
	$\Delta E_{m}$ is of the magnetic one.
	$\Delta E = \Delta E_{e} + \Delta E_{m}$.
	}
	\begin{ruledtabular}
	\begin{tabular}{cccc}
	$S$	&$\Delta E_{e}$ (MeV)	&$\Delta E_{m}$ (MeV)	&$\Delta E$ (MeV)\\
	\hline
	 0	& -73.0		& -222.8	& -296\\
	-1	& -58.2		& -184.7	& -243\\
	-2	& -54.5		& -161.7	& -216\\
	-3	& -53.9		& -160.7	& -215\\
	\end{tabular}
	\end{ruledtabular}
\label{tb:casimir_energy}
\end{table}

We summaries the our results in Table~\ref{tb:rot_band}.
Since $\Delta E$ take negative values,
the calculated masses without $\Delta E$ should be larger than the empirical values.
In Ref.~\cite{rf:Akiyama04}, the current s quark mass $m_{s} = 200$ MeV
is an adequate value so that the masses without the Casimir energies reproduce
the empirical ones for the hyperons.
Larger $m_{s}$ in this paper pushes up the hyperon masses and cancels out the the Casimir energy.
However we can not find the moderate values of $m_{s}$ to reproduce the hyperon masses accurately.
It is because that the soliton breaks for the large $m_{s}$
due to the lack of the confinement in this model. 
For the dynamically generated quark mass $M = 400$ MeV
and the cutoff parameter $\Lambda = 700$ MeV,
there is the critical value in the vicinity of $m_{s} = 210$ MeV.
As a result, the error due to our approach is at most 15\% for the octet and decouplet
baryon masses.
\begin{table}
	\caption{The masses of the octet and decuplet baryons.
	$S$ is the strangeness of baryon.
	$E$ is the contribution from the classical soliton and the collective motion.
	$\Delta E$ is the Casimir energy.
	Expt. is the empirical value.
	}
	\begin{ruledtabular}
	\begin{tabular}{ccccc}
	$S$ & Particle & $E$ (MeV) & $E+\Delta E$ (MeV) & Expt. (MeV)\\
	\hline
	0	& $N$			& 1410 & 1114 &  939\\
		& $\Delta$		& 1648 & 1352 & 1232\\
	\hline
	-1	& $\Lambda$		& 1224 &  981 & 1116\\
		& $\Sigma$		& 1249 & 1006 & 1195\\
		& $\Sigma^{*}$	& 1476 & 1233 & 1384\\
	\hline
	-2	& $\Xi$			& 1405 & 1189 & 1318\\
		& $\Xi^{*}$		& 1701 & 1485 & 1530\\
	\hline
	-3	& $\Omega$		& 1788 & 1573 & 1672\\
	\end{tabular}
	\end{ruledtabular}
\label{tb:rot_band}
\end{table}


\section{Summary\label{sec:summary}}
We have studied the Casimir energies for the octet and decuplet baryons
in the SU(3) chiral quark soliton model.
They are the quantum corrections due to the meson fluctuations around the soliton.
Especially we concentrate on the zero mode contributions
in the baryon number $B = 1$ sector.
Since the soliton takes the non-hedgehog shape
for the hyperons ($\Lambda$,$\Sigma^{(*)}$,$\Xi^{(*)}$,$\Omega$) in our approach,
we develop a method to treat the fluctuations around the non-hedgehog soliton.

Since the fluctuations are some bound states of the quarks in this model,
those equations of motion are the Bethe-Salpeter equations.
We adopt an ans\"{a}tz to reduce the degrees of freedom of the fluctuations.
As a result, the fluctuations are separated into the electric and magnetic parts.
In addition, each of them can be distinguished with the spherical indices
due to the axial symmetry of the soliton.
Thus we have six independent zero modes.

The fluctuations for hyperons shrink toward the center of the soliton
compared with the cases of the non-strange baryons.
This feature is because that the quarks constituting the soliton become heavy
for the hyperons due to the flavor rotation into strange direction.

The regularized Casimir energy is defined through the overlap integrals
between the fluctuations in the $B = 0, 1$ sectors.
The integrals can be regarded as the functions of the energy eigenvalue
of the fluctuations in the $B = 0$ sector.
Then the integrals for the hyperons are shifted to the higher energy region
than ones for the non-strange baryons. 
It is consistent with the shrink of the fluctuations for hyperons.

According to the quark picture of the fluctuations,
we employ $2\times$(the lightest constituent quark mass) as a cutoff for the Casimir energy.
Then the fluctuations for the hyperons are suppressed in comparison
with the cases of the non-strange baryons.
It is reasonable from the physical point of view.
Our approach reproduces the masses for the octet and decuplet baryons
within 15 \% error at the most.

\appendix
\section{Cutoff function\label{sec:cutoff}}
We use the Schwinger proper time regularization~\cite{rf:Schwinger51,rf:Ebert86,rf:Diakonov88,rf:Weigel95}
in this paper.
The cutoff function for the local density is given by
\begin{equation}
	\rho^{R}\left(\epsilon,\Lambda\right) = N_{c} \eta^{val}_{\epsilon}
		+ N_{c} \sgn\left(\epsilon\right) {\cal N}^{R}\left(\epsilon, \Lambda\right),
\end{equation}
where $\eta^{val}_{\epsilon}$ is the occupation number of valence quark in the level $\epsilon$ and
\begin{equation}
	\sgn\left(\epsilon\right) {\cal N}^{R}\left(\epsilon, \Lambda\right) =
		-\int^{\infty}_{1/\Lambda^{2}} d\rho \frac{\epsilon}{\sqrt{4\pi\rho}} e^{-\rho \epsilon^{2}}  =
		-\half \sgn\left(\epsilon\right) \erfc\left(\abs{\epsilon/\Lambda}\right).
\end{equation}

The cutoff function for the bi-local density is given by
\begin{equation}
	\rho_{\Gamma_{2}}\left(\epsilon_{a},\epsilon_{b},\omega,\Lambda\right) =
		N_{c} \frac{\eta^{val}_{\epsilon_{a}}-\eta^{val}_{\epsilon_{b}}}{\omega+\epsilon_{b}-\epsilon_{a}}
		+ \half N_{c} f_{\Gamma_{2}}\left(\epsilon_{a},\epsilon_{b},\omega,\Lambda\right),
\end{equation}
where
\begin{equation}
	f_{\Gamma_{2}}\left(\epsilon_{a}, \epsilon_{b}, \omega, \Lambda \right) =
		\int^{\infty}_{1/\Lambda^{2}} d\rho \sqrt{\frac{\rho}{4\pi}}
		\left\{
			\frac{e^{-\rho\epsilon_{a}^{2}}+e^{-\rho\epsilon_{b}^{2}}}{\rho}
			+\left[
				\omega^{2}-\left(\epsilon_{a}+\epsilon_{b}\right)^{2}
			\right] R_{0}(\rho,\omega,\epsilon_{a},\epsilon_{b})
		\right\},
\end{equation}
and
\begin{equation}
	R_{0}(\rho,\omega,\epsilon_{a},\epsilon_{b}) = \int^{1}_{0} d\alpha
		\exp\left(-\rho\left[(1-\alpha) \epsilon_{a}^{2}+\alpha  \epsilon_{b}^{2}-\alpha(1-\alpha) \omega^{2}\right] \right).
\end{equation}

\section{Bethe-Salpeter kernel\label{sec:bs_kernel}}
By using the ans\"{a}tz (\ref{eq:ze_nh}) in Eq.~(\ref{eq:sz_reduce}) and integrating over $\varphi$,
we obtain the action Eq.~(\ref{eq:sz_e}) for the electric fluctuations.
The local kernel $\Phi^{(1)E}_{m}$ becomes a $2\times2$ diagonal matrix,
\begin{eqnarray}
	\Phi^{(1)E}_{m}\left(r,\theta\right) &=&
		M \left(
			\begin{array}{cc}
				\left(-1\right)^{m} \hat{\Lambda}_{-m}\hat{\Lambda}_{m} & 0\\
				0 &	1-\left(-1\right)^{m}\hat{\Lambda}_{-m}\hat{\Lambda}_{m}			
			\end{array}
		\right) \nonumber\\
		&&\times\left[S \cos F + ({\bf P} \cdot \hat{\bf \Lambda}) \sin F\right],
\end{eqnarray}
where $S$ and $P_{m}$ are defined at Eqs.~(\ref{eq:scalar_density}),(\ref{eq:pscalar_density}).
Under the situation Eq.~(\ref{eq:k3_symmetry}), the quantities $S$, $F$, $(\hat{\Lambda}_{-m}\hat{\Lambda}_{m})$,
and $({\bf P} \cdot \hat{\bf \Lambda})$ are independent of $\varphi$ from the first.

The bilocal kernel $\Phi^{(2)E}_{m}$ is given by 
\begin{eqnarray}
	\Phi^{(2)E}_{m}\left(r,\theta,r',\theta',\omega\right) &=&
		\left(-1\right)^{m} M^{2} \sum_{a,b}
		\rho_{\Gamma_{2}}\left(\epsilon_{a},\epsilon_{b},\omega,\Lambda\right) \nonumber\\
		&&\times\left(
			\begin{array}{cc}
				L^{(a,b)}_{-m}\left(r,\theta\right) L^{(b,a)}_{m}\left(r',\theta'\right) &
				L^{(a,b)}_{-m}\left(r,\theta\right) T^{(b,a)}_{m}\left(r',\theta'\right) \\
				T^{(a,b)}_{-m}\left(r,\theta\right) L^{(b,a)}_{m}\left(r',\theta'\right) &
				T^{(a,b)}_{-m}\left(r,\theta\right) T^{(b,a)}_{m}\left(r',\theta'\right)				
			\end{array}
		\right),
\end{eqnarray}
where
\begin{eqnarray}
	L^{(a,b)}_{m}\left(r,\theta\right) &=& \delta_{M_{a},M_{b}+m}
		\hat{\Lambda}_{m}\left(r,\theta,0\right)
		\left[
			-S^{(a,b)}\left(r,\theta,0\right) \sin F
			+({\bf P}^{(a,b)} \cdot \hat{\bf \Lambda})\left(r,\theta,0\right) \cos F
		\right],\\
	T^{(a,b)}_{m}\left(r,\theta\right) &=& \delta_{M_{a},M_{b}+m} \sum_{n}
		\left[
			g_{m}{}^{n}-\hat{\Lambda}_{m}\left(r,\theta,0\right) \hat{\Lambda}^{n}\left(r,\theta,0\right)
		\right] P^{(a,b)}_{n}\left(r,\theta,0\right),
\end{eqnarray}
and $S^{(a,b)}$ and $P^{(a,b)}_{m}$ are defined at Eqs.~(\ref{eq:scalar}),(\ref{eq:pscalar}).
The integral over $\varphi$ yields the conservation of grand spin $\delta_{M_{a},M_{b}+m}$
and reduction of computing load.

For the magnetic fluctuations, by using Eq.~(\ref{eq:zm_nh}),
we obtain the action (\ref{eq:sz_m}).
The local kernel is given by
\begin{equation}
	\Phi^{(1)M}_{m}\left(r,\theta\right) =
		 M \left[1-\left(-1\right)^{m}\hat{\Lambda}_{-m}\hat{\Lambda}_{m}\right]
		\left[S \cos F + ({\bf P} \cdot \hat{\bf \Lambda}) \sin F\right].
\end{equation}
And the bilocal kernel is given by
\begin{equation}
	\Phi^{(2)M}_{m}\left(r,\theta,r',\theta',\omega\right) =
		-\left(-1\right)^{m} M^{2} \sum_{a,b}
		\rho_{\Gamma_{2}}\left(\epsilon_{a},\epsilon_{b},\omega,\Lambda\right)
		C^{(a,b)}_{-m}\left(r,\theta\right) C^{(b,a)}_{m}\left(r',\theta'\right),
\end{equation}
where
\begin{equation}
	C^{(a,b)}_{m}\left(r,\theta\right) = \delta_{M_{a},M_{b}+m} \sum_{m_{1},m_{2}}
		\frac{1}{i} \varepsilon_{m}{}^{m_{1}m_{2}} \hat{\Lambda}_{m_{1}}\left(r,\theta,0\right)
		P^{(a,b)}_{m_{2}}\left(r,\theta,0\right).
\end{equation}
Similarly, there is the conservation of grand spin $\delta_{M_{a},M_{b}+m}$.

\section{Wave functions for $B = 0$ sector}\label{sec:b0_wf}
We define the normalized radial function:
\begin{equation}
	u_{l}(k r) = \frac{\sqrt{2}}{\abs{j_{l+1}(k R)}} j_{l}(k r),
\end{equation}
where $l$ is the angular momentum, $j$ is the spherical Bessel function,
$k$ is the momentum, and $R$ is the radius of the large spherical box.
The momentum $k$ is determined by the boundary condition
corresponding to Eq.~(\ref{eq:boundary_cond}),
\begin{equation}
	j_{l}(k_{l} R) = 0.
\end{equation}

The normalized wave functions Eq.~(\ref{eq:wave_func}) for the electric fluctuations become
\begin{equation}
	\phi^{(0)m}(r,\theta,\omega) =
	\left(
		\begin{array}{c}
		\sqrt{(-1)^{m} \hat{x}_{-m} \hat{x}_{m}} \zeta^{Lm}(r,\omega)\\
		\sqrt{1-(-1)^{m} \hat{x}_{-m} \hat{x}_{m}} \zeta^{Tm}(r,\omega)\\
		\end {array}
	\right),
\end{equation}
where $\zeta^{Lm}$ and $\zeta^{Tm}$ are the radial part of the wave function,
$\hat{x}_{m} = x_{m}/r$, and $\hat{x}_{-m} \hat{x}_{m}$ is independent of $\varphi$.
For the S-wave fluctuations, the radial parts are given by
\begin{eqnarray}
	\zeta^{Lm}(r,\omega) &=& \frac{1}{\sqrt{4\pi}} u_{0}(k_{0} r),\\
	\zeta^{Tm}(r,\omega) &=& \frac{1}{\sqrt{4\pi}} u_{0}(k_{0} r),
\end{eqnarray}
where $\omega = \sqrt{W_{m}^{2} + k_{0}^{2}}$.
For the D-wave fluctuations,
\begin{eqnarray}
	\zeta^{Lm}(r,\omega) &=& \frac{1}{\sqrt{2\pi}} u_{2}(k_{2} r),\\
	\zeta^{Tm}(r,\omega) &=& -\frac{1}{\sqrt{8\pi}} u_{2}(k_{2} r), 
\end{eqnarray}
where $\omega = \sqrt{W_{m}^{2} + k_{2}^{2}}$.

The normalized wave functions for the magnetic fluctuations become
\begin{equation}
	\phi^{(0)m}(r,\theta,\omega) = \sqrt{1-(-1)^{m} \hat{x}_{-m}\hat{x}_{m}} \zeta^{Mm}(r,\omega),
\end{equation}
where
\begin{equation}
	\zeta^{Mm}(r,\omega) = \sqrt{\frac{3}{8\pi}} u_{1}(k_{1} r)
\end{equation}
with $\omega = \sqrt{W_{m}^{2} + k_{1}^{2}}$.


\end{document}